\documentclass[aps,prl,twocolumn,nofootinbib,twocolumn,floatfix,showpacs,superscriptaddress]{revtex4-2}
\usepackage{amsmath,graphicx,epsfig,amsthm,color,bm}
\usepackage{pifont,bbding,amssymb,soul,mathrsfs,braket}
\usepackage{dcolumn}% Align table columns on decimal point
\usepackage{siunitx}
\usepackage[hyperindex,breaklinks,colorlinks=true,citecolor=blue,urlcolor=blue]{hyperref}
\usepackage{subfigure}

\usepackage{threeparttable}
\usepackage{booktabs}

\begin{document}
\title{High-efficiency On-chip Quantum Photon Source in Modal Phase-matched Lithium Niobate Nanowaveguide}
\author{Xiao-Xu Fang}
\email{These authors contributed equally to this work.}
\affiliation{School of Physics, State Key Laboratory of Crystal Materials, Shandong University, Jinan 250100, China}
\affiliation{Shenzhen Research Institute of Shandong University, Shenzhen 518057, China}
\author{Hao-Yang Du}
\email{These authors contributed equally to this work.}
\affiliation{School of Physics, State Key Laboratory of Crystal Materials, Shandong University, Jinan 250100, China}
\author{Xiuquan Zhang}
\affiliation{Key Laboratory of Laser \& Infrared System, Ministry of Education Shandong University, Qingdao 266000, China.}
\author{Lei Wang}
\affiliation{School of Physics, State Key Laboratory of Crystal Materials, Shandong University, Jinan 250100, China}
\author{Feng Chen}
\email{drfchen@sdu.edu.cn}
\affiliation{School of Physics, State Key Laboratory of Crystal Materials, Shandong University, Jinan 250100, China}

\author{He Lu}
\email{luhe@sdu.edu.cn}
\affiliation{School of Physics, State Key Laboratory of Crystal Materials, Shandong University, Jinan 250100, China}
\affiliation{Shenzhen Research Institute of Shandong University, Shenzhen 518057, China}

\begin{abstract}
Thin-film lithium niobate on insulator~(LNOI) emerges as a promising platform for integrated quantum photon source, enabling scalable on-chip quantum information processing. The most popular technique to overcome the phase mismatching between interacting waves in waveguide is periodic poling, which is intrinsically sensitive to poling uniformity. Here, we report an alternative strategy to offset the phase mismatching of spontaneous parametric down-conversion~(SPDC) process, so-called modal phase matching, in a straight waveguide fabricated on a dual-layer LNOI. The dual-layer LNOI consists of two 300~nm lithium niobates with opposite directions, which significantly enhances the spatial overlap between fundamental and high-order modes and thus enables efficient SPDC. This dual-layer waveguide generates photon pairs with pair generation rate of 41.77~GHz/mW, which exhibits excellent signal-to-noise performance with coincidence-to-accidental ratio up to 58298$\pm$1297. Moreover, we observe a heralded single-photon source with second-order autocorrelation $g_{H}^{(2)}(0)<0.2$ and heralded rate exceeding 100~kHz. Our results provide an experiment-friendly approach for efficient generation of quantum photon sources and benefit the on-chip quantum information processing based on LNOI.
\end{abstract}
\maketitle

{\noindent\bf\em Introduction.}---
Lithium niobate~(LN) is a well-established material for frequency conversion, both in classical and quantum realm, owing to its strong second-order nonlinear susceptibility~($d_{33}$=-34.4~pm/V) and large transparency window spanning from ultraviolet to mid-infrared wavelengths~(0.35-5.2~$\mu$m)~\cite{1985_weis_Appl.Phys.A_Lithium}. The rapid development of thin-film LN on insulator~(LNOI) together with the structuring approach based on lithography and dry etching, significantly enhances the efficiency of frequency conversion. Specifically, spontaneous parametric down-conversion~(SPDC) process~\cite{Burnham1970PRL} is one of the most prominent technologies to generate correlated photon pairs, which is the building block in quantum information science with photonic system~\cite{Walther2005Nature,Yao2012Nature,Lu2016PRL,Zhong2018PRL}. Waveguide on LNOI further enhances the efficiency of SPDC, benefiting from the strong confinement of interacting waves. The key ingredient to achieve high-efficient SPDC on waveguide is to offset the phase mismatching between interacting waves, which is induced by the difference between effective refractive indices of interacting waves. The second-order nonlinear tensor $\chi^{(2)}$ can be periodically controlled by electric-field poling, enabling quasi-phase matching~(QPM) to overcome the phase mismatching. QPM with periodical poling has been widely adopted for efficient SPDC on LNOI~\cite{zhao2020_Phys.Rev.Lett._High, javid2021_Phys.Rev.Lett._Ultrabroadband, xue2021_Phys.Rev.Applied_Ultrabright, henry2023_Opt.Express_Correlated, zhang2023_Optica_Scalable}. However, periodical poling is intrinsically suffered from the non-uniformity of the duty cycle, which retards its conversion efficiency compared to theoretical predictions~\cite{Helmfrid:91,chang2016_Optica_Thin,wang2018_Optica_Ultrahigh-efficiency,wu2022_Opt.Lett.OL_Broadband,zhang2023_Optica_Scalable}.  

Alternative strategies to overcome the phase mismatching have been proposed and demonstrated~\cite{wu2022_Opt.Lett.OL_Broadband, tang2023_Opt.Lett._Broadband,chuanyizhu2017_Chin.Opt.Lett._Multiple-mode, chen2018_OSAContinuum_Modal,luo2018_Optica_Highly,luo2019_Phys.Rev.Applied_Optical,luo2019_Laser&PhotonicsReviews_Semi,wang2020_ResultsinPhysics_Type,wang2021_Laser&PhotonicsReviews_Efficient,du2023_Opt.Lett._Tunable,wang2023_Photonics_Broadband, hu2023_Opt.Eng._Design}. Birefringent phase matching~(BPM) is strict phase matching and easy to fabricate, but is restricted to frequency conversion with the smaller nonlinear susceptibility~($d_{31}$ = -4.35 pm/V) of LN~\cite{lu2022_Opt.Lett._Highly, tang2023_Opt.Lett._Broadband}. 
Modal phase matching~(MPM) is another strict phase matching technique utilizing $d_{33}$ for wavelength conversion, in which the effective refractive indices of different wavelengths are equal at fundamental and higher-order modes. In contrast to QPM, nonlinear devices based on MPM are easy to fabricate without post-processing steps. However, the weak modal overlap between interacting waves limits its conversion efficiency~\cite{wang2017_Opt.ExpressOE_Second, chen2018_OSAContinuum_Modal,luo2018_Optica_Highly}. The issue of weak modal overlap can be addressed with the dual-layer device. A heterogeneously integrated waveguide, i.e., TiO$_2$~$(\chi=0)$ on LN~$(\chi \ne 0)$, was reported to improve the modal overlap, enabling the observed conversion efficiency in second harmonic generation~(SHG) up to 650$\% \text{W}^{-1} \text{cm}^{-2}$~\cite{luo2019_Laser&PhotonicsReviews_Semi}. There is no second-order nonlinear susceptibility in TiO$_2$, so that such kind of device is so-called semi-nonlinear device. Dual-layer LNOI is capable to further improve the conversion efficiency, and has been employed to realize high-efficient sum-frequency generation~(SFG)~\cite{du2023_Opt.Lett._Tunable} and  SHG~\cite{wang2021_Laser&PhotonicsReviews_Efficient,li2023_2023AsiaCommun.PhotonicsConf.Int.PhotonicsOptoelectron.Meet.ACPPOEM_Exceptionally}, with conversion efficiency up to $9700\% \text{W}^{-1} \text{cm}^{-2}$~\cite{li2023_2023AsiaCommun.PhotonicsConf.Int.PhotonicsOptoelectron.Meet.ACPPOEM_Exceptionally}.

Although MPM has shown great advantages in efficient wavelength conversion, the MPM-enabled quantum photon source is still out of research. In this work, we report an efficient SPDC source enabled by MPM on a dual-layer LNOI waveguide. We obtain a two-photon source with pair generation rate~(PGR) of 41.77~GHz/mW and coincidence-to-accidental ratio up to 58298$\pm1297$, and a heralded single photon source with second-order autocorrelation less than 0.2 and rate exceeding 100~kHz. The results of MPM-enabled SPDC are comparable to SPDC source with periodical poling, while alleviates the experimental complexity in fabrication.

\begin{figure*}[htb]%4.25
\includegraphics[width=\linewidth]{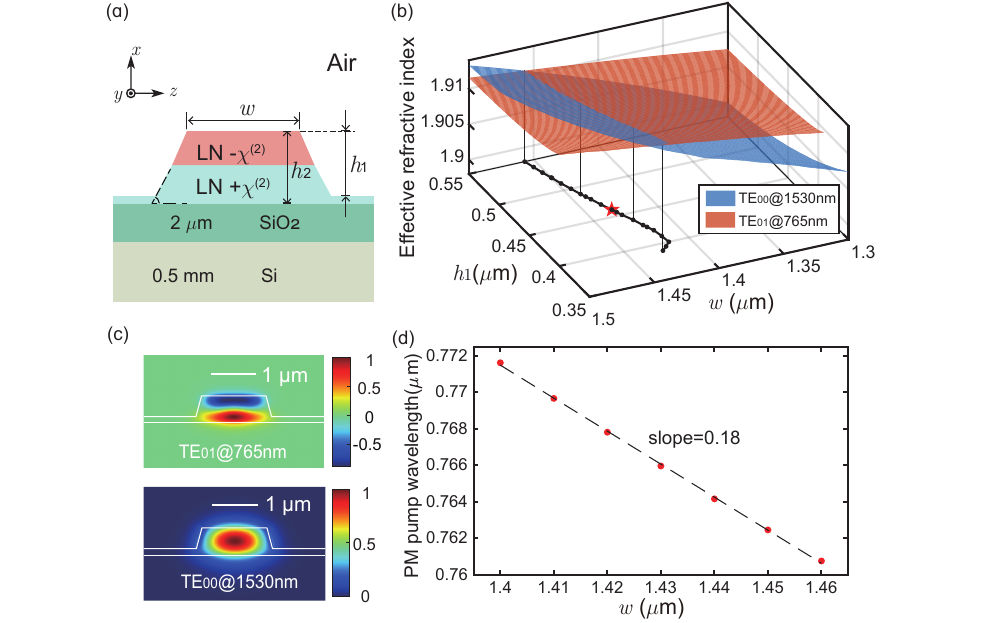} 
\caption{(a) The ridge waveguide is designed and fabricated on a 600~nm thick dual-layer $x$-cut LNOI wafer consisting of two 300~nm-thick LN layers with opposite $z$ directions. The waveguide is designed with sidewall angle of $\theta=75^{\circ}$, top width of $w$, and etch depth of $h_1$. (b) Simulation results of the effective refractive index $n_\text{eff}$ of TE$_{00}$ at 1530~nm~(blue landscape) and TE$_{01}$ at 765~nm~(red landscape), with changes of $w$ and $h_1$. (c) The electric field profiles along $z$ direction of TE$_{01}$ at 765~nm and TE$_{00}$ at 1530~nm respectively. (d) Simulation results of wavelength of pump light satisfying MPM with changes of width $w$. } 
\label{Fig:design}
\end{figure*}

{\noindent\bf\em Design and Simulation.}---
In SPDC, a pump photon~($p$) is spontaneously split into a pair of signal~($s$) and idler~($i$) photons. The high-efficiency SPDC requires energy and momentum conservation, i.e., $\omega_p=\omega_s+\omega_i$ and $\Delta k = k_s+ k_i - k_p=0$ where $\omega$ is frequency, $k=\frac{2\pi}{\lambda}\cdot n_\text{eff}$ is the wavevector and $n_\text{eff}$ is the effective refractive index. In the degenerate case $\omega_s=\omega_i$, the phase matching $\Delta k = 0$ implies $n_\text{eff}(\omega_p)=n_\text{eff}(\omega_{s(i)})$. The effective refractive indices satisfy $n_{\mathrm{eff}}(\omega_p) > n_{\text{eff}}(\omega_s)$ if the pump and signal lights travel in the same spatial modes, thus the phase matching cannot be achieved. In MPM, the pump and signal lights are engineered to travel in two specific modes of waveguide, satisfying $n_\text{eff}(\omega_p) = n_\text{eff}(\omega_s)$. As shown in Figure~\ref{Fig:design}~(a), the sample of LNOI consists of two $x$-cut two 300~nm-thick thin-film LN with opposite $z$ directions, which is obtained by bonding two LNOI wafers~\cite{wang2021_Laser&PhotonicsReviews_Efficient}. A ribbed straight waveguide is then directly fabricated on such a dual-layer LNOI using E-beam lithography, lift-off processing and dry etching, similar to the technique reported in~\cite{du2024_Opt.Lett._High-efficiency}. The MPM is taken place between the fundamental quasi-transverse-electric mode TE$_{00}$ at 1530~nm (signal light $\omega_s$) and the higher-order quasi-transverse-electric mode TE$_{01}$ at 765~nm (pump light $\omega_p$). The effective refractive indices $n_\text{eff}$ are highly related to the top width $w$ and etch depth $h_1$ of the waveguide. We numerically simulate the refractive indices of modes TE$_{00}$ and TE$_{01}$ with different $w$ and $h_1$, and the results are shown with the blue and red landscapes in Figure~\ref{Fig:design}~(b). The intersection of two landscapes indicates the phase-matching condition $n_\text{eff}(\omega_p)=n_\text{eff}(\omega_s)$, in which the etch depth $h_1>390$~nm guarantees the existence of mode TE$_{\text{01}}$. By fixing $h_1=460$~nm, the top width $w \approx 1.43~\mu$m according to the phase matching condition, which is marked by a red star in Figure~\ref{Fig:design}~(b). 

The mode profiles of TE$_{01}$ at 765 nm and TE$_{00}$ at 1530~nm with $h_1=460$~nm and $w=1.43~\mu$m are shown in Figure~\ref{Fig:design}~(c). The PGR of SPDC process, which indicates the coincidence of photon pairs generated per second, can be shown to have the following dependence~\cite{wang2021_AppliedPhysicsReviews_Integrated, cheng2019_Opt.Express_Design}
\begin{equation}
\label{eq:eff}
\text{PGR}\propto L^2P_p\frac{{d_{\mathrm{eff}}^{2}}\zeta ^2}{A_{\mathrm{eff}}}\sin\mathrm{c}^2(\frac{\Delta kL}{2}),
\end{equation}
where $L$ is the waveguide length, $P_p$ is the pump power, $d_{\mathrm{eff}}=d_{33}$ and $A_{\mathrm{eff}}$ is the effective mode area. $\zeta$ is the modal overlap factor~\cite{wang2021_Laser&PhotonicsReviews_Efficient}
\begin{widetext}
\begin{equation}
\label{eq:overlap}
\zeta =\frac{\iint{d_{\text{Nor}}}(x,z)\cdot E_{\text{TE}_{00}}(x,z)^{\ast}E_{\text{TE}_{00}}(x,z)^{\ast}E_{\text{TE}_{01}}(x,z)\text{d}x\text{d}z}{\left| \iint{\left| E_{\text{TE}_{00}}(x,z) \right|^2E_{\text{TE}_{00}}(x,z)\text{d}x\text{d}z} \right|^{2/3}\left| \iint{\left| E_{\text{TE}_{01}}(x,z)\right|^2d_{\text{Nor}}(x,z)E_{\text{TE}_{01}}(x,z)\text{d}x\text{d}z} \right|^{1/3}},
\end{equation}
\end{widetext}
where $d_\text{Nor}(x,z) = \pm 1$ is the normalized space profile of the nonlinear susceptibility $\chi^{(2)}$ and $E(x,z)$ is electric field of transverse waveguide mode along $z$ direction. We simulate the value $\zeta$ of single-layer and dual-layer LN waveguides respectively according to Eq.~\ref{eq:overlap}. For single-layer case, $d_\text{Nor}(x,z)=1$ and $\zeta=0.21$. For dual-layer case, $d_\text{Nor}(x,z) =  1$ for the bottom layer and $d_\text{Nor}(x,z) =  -1$ for the top layer as shown in Figure~\ref{Fig:design}~(a), and modal overlap factor $\zeta=0.81$. Accordingly, the conversion efficiency is expected to be enhanced by approximately 16 times in dual-layer device.

\begin{figure}[htb]%4.25
\centering
\includegraphics[width=\linewidth]{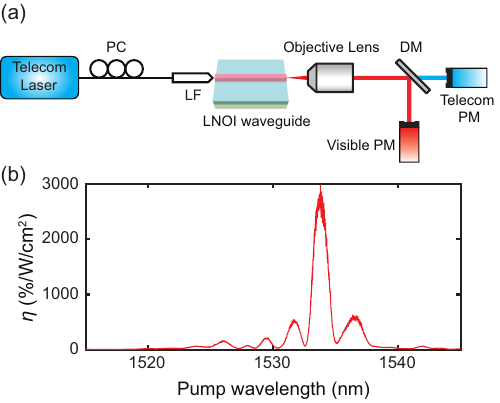} 
\caption{(a) The experimental setup to characterize the SHG of LNOI waveguide. (b) Measured normalized SHG efficiency. PC: polarization controller; LF: lensed fiber; DM: dichroic mirror; PM: power meter.} 
\label{Fig:shg}
\end{figure}

{\noindent\bf\em Experimental Results of SHG.}---
We first characterize the SHG of the fabricated LNOI waveguide, which is the most fundamental three-wave mixing process to evaluate the nonlinearity. We characterized the efficiency of SHG with the setup shown in Figure~\ref{Fig:shg}(a). The first-harmonic~(FH) light from a continuous-wave~(CW) laser in telecom band is coupled into the LN waveguide via a lensed fiber~(LF), where a polarization controller~(PC) is used to ensure the TE polarization of FH light. The LNOI chip is mounted on a thermometric cooler set at $23^{\circ}$. The second-harmonic~(SH) light, along with the FH light, is coupled out with an objective lens and then is separated by a dichroic mirror. The power of SH light $P_\text{SH}$ and FH light $P_\text{FH}$ are measured by a visible power meter~(PM) and a telecom PM, respectively. The coupling losses~(waveguide to PM) of FH and SH light are 2.6~dB and 2.3~dB. The propagation loss of FH light is estimated to be 1.8~dB/cm using the Fabry-Perot method~\cite{chen2020_Opt.Lett._Efficient}. The roughness of waveguide surface induces Rayleigh scattering of the propagation light, which consequently introduces propagation proportional to $1/\lambda^2$ \cite{luo2019_Laser&PhotonicsReviews_Semi}. The conversion efficiency $\eta$ of SHG is calculated by $\eta=\frac{P_\text{SH}/T_\text{SH}}{(P_\text{FH}L/T_\text{FH})^{2}}$, where $L=5.2$~mm is the length of LNOI waveguide, $T_\text{SH}=35\%$ and $T_\text{FH}=30\%$ are the transmissivity of SH and FH light respectively. The results of $\eta$ are shown in~Figure~\ref{Fig:shg}(b), and we observe the maximal normalized SHG efficiency of 2976\%/W/cm$^2$ at FH wavelength of 1536.8~nm. The experimental result of conversion efficiency of SHG is smaller than simulation value, i.e., 10419\%/W/cm$^2$. The discrepancy between experiment and simulation is mainly attributed to the non-uniformity of the width of the fabricated waveguide and the thickness of the dual-layer LNOI, which introduces the phase mismatching~\cite{du2023_Opt.Express_Highly,kuo2022_Opt.Lett._Noncritical}.

\begin{figure*}[h!tbp]%4.25
 \centering
 	\includegraphics[width=0.8\linewidth]{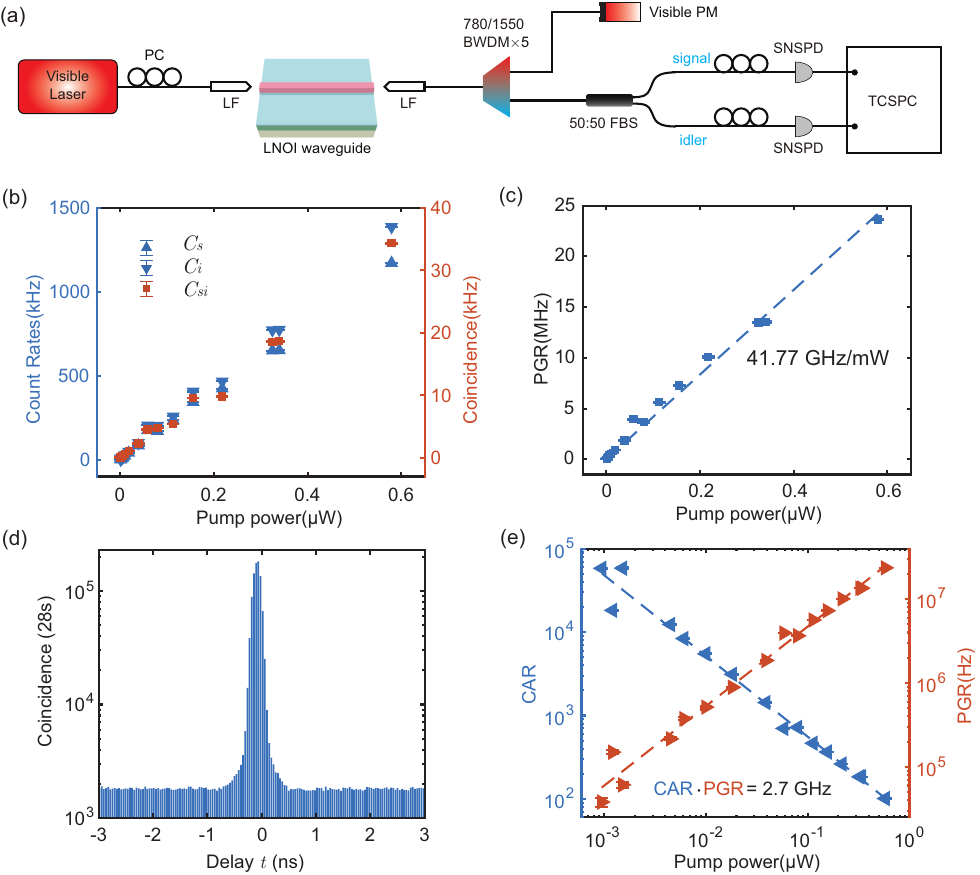} 
 	\caption{(a) The experimental setup to characterize the generated photon pairs from LNOI waveguide. (b) The count rates of signal photon, idler photon and two-photon coincidences under different pump power. (c) PGRs under different pump power. (d) Coincidence at different delay $t$ by accumulating 28s. (e) CARs and PGRs under different pump power. The error bars are standard deviations estimated by assuming that the collected data are with Poisson statistics. 780/1550 BWDM: 780/1550~nm band wavelength division multiplexer; 50:50 FBS: 50:50 fibered beamsplitter; SNSPD: superconducting nanowire single-photon detectors; TCSPC: time-correlated single-photon counting system.} 
 	\label{Fig:SPDC}
\end{figure*}

{\noindent\bf\em Experimental Results of SPDC.}---
The results of SHG indicate the capability of fabricated LNOI waveguide for SPDC. As shown in~Figure~\ref{Fig:SPDC}~(a), the pump light from a CW laser~(New Port, TLB 6700) set at 768.4~nm is coupled into the LNOI waveguide via a LF, where the pump light is adjusted to TE polarization by a PC for type-0 SPDC. The observation of two-photon coincidence indicates that TE$_{01}$ mode is coupled into the waveguide as only TE$_{01}$ mode is able to excite SPDC in our waveguide. Then, we slightly adjust the position of LF to obtain the maximal two-photon coincidence. The generated signal and idler photons are coupled out from the chip with a second LF, and the pump light is filtered out by five 780/1550~nm band wavelength division multiplexers~(BWDM). The power intensity of pump light is measured by a visible PM. To measure the brightness of photon-pair generation from the LNOI waveguide, we use a 50:50 fibered beam splitter~(FBS) to separate the signal and idler photons, which are detected by two superconducting nanowire single-photon detectors~(SNSPDs) located in a cryostat~(Quantum Opus, Opus One) and recorded by a time-correlated single-photon counting~(TCSPC) system~(Swabian Instruments, Time Tagger 20) with coincidence window of 1~ns. We denote the counter rates of signal and idler photons as $C_s$ and $C_i$ respectively, and the coincidence as $C_{si}$. The experimental results of $C_s$, $C_i$ and $C_{si}$ at different pump powers are shown in ~Figure~\ref{Fig:SPDC}(b), according to which we calculate the PGR by $C_iC_s/2C_{si}$. The factor of 2 in the denominator is induced by the FBS in the experimental configuration for detection. Note that the pump power here is the estimated power intensity coupled into TE$_{01}$. To this end, we measure the coupling loss between LF and TE$_{00}$ mode of waveguide, which is 6~dB and agrees well with simulation. Then, we simulate the coupling loss between LF and TE$_{01}$ mode, which is 20.7~dB and is used to estimate the power intensity coupled into mode TE$_{01}$. The results of PGR are shown in Figure~\ref{Fig:SPDC}~(c). By linear fitting of the PGRs, we obtain a slope of 41.77~GHz/mW.

We also measure the temporal correlation of the generated photons, which is determined by the coincidence-to-accidental ratio~(CAR). The signal-idler coincidence accumulated in 28~s with different time delay as shown in~Figure~\ref{Fig:SPDC}~(d). CAR is calculated by $\text{CAR}=\max[g_{si}^{(2)}(t)]-1$ with $g_{si}^{(2)}(t)=C_{si}(t)/C_{si}(\infty)$, where $C_{si}(t)$ is the coincidence at time delay $t$ between signal and idler photons and $C_{si}(\infty)$ is the coincidence at time delay far from $t=0$. The CARs under different pump powers are shown with blue triangles in~Figure~\ref{Fig:SPDC}~(e). In contrast to the PGR, the CAR is inversely proportional to the power of the pump light. In our experiment, the highest CAR of $58298 \pm 1297$ is achieved at pump power of 20.8~nW, and the corresponding PGR is $61 \pm 5$~kHz. The product of CAR and PGR is independent of the pump power, and we obtain $\text{CAR}\times \text{PGR}\approx2.7$~GHz.

\begin{figure}[h!]%4.25
 \centering
 	\includegraphics[width=\linewidth]{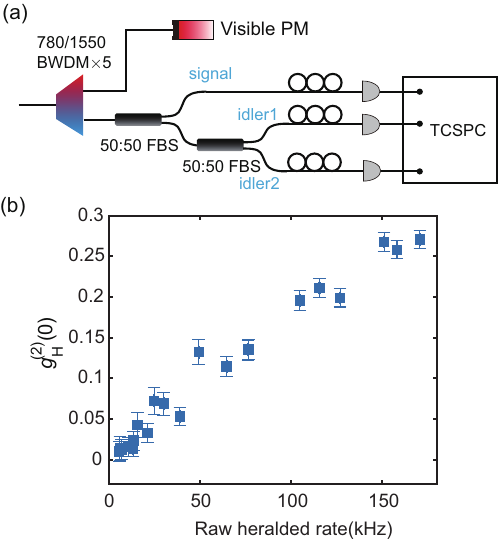} 
 	\caption{(a) The setup to measure the second-order autocorrelation function of heralded single photon. (b) The second-order autocorrelation function $g_{\text{H}}^{(2)}(0)$ of the heralded photon with various pump power. The error bars are the statistic errors obtained with Monte Carlo simulation by assuming the collected counts are with Poisson distribution.} 
 	\label{Fig:g2}
\end{figure}

Finally, we characterize the property of heralded single-photon source~(HSPS) based on the SPDC, in which the detection of one photon is used to herald the emission of the other one. As shown in Figure~\ref{Fig:g2}~(a), the signal photon is detected to herald the emission of the idler photon, which is characterized by the Hanbury Brown and Twiss effect~\cite{brown_correlation_1956}. The idler photon is split by a 50:50 FBS and detected by two detectors $i_1$ and $i_2$. The nonclassical antibunching of HSPS is characterized by second-order autocorrelation function of idler photon with zero time delay, i.e., $g_{H}^{(2)}(0)=C_{si_1i_2}C_{s}/2C_{si_1}C_{si_2}$, where $C_{si_1i_2}$ is the three-folder coincidence between detectors $s$, $i_1$ and $i_2$. The results of $g_{H}^{(2)}(0)$ are shown in~Figure~\ref{Fig:g2}~(c). At pump power of 0.25~$\mu$W, we observe the HSPS with raw heralded rate $C_{si_1}+C_{si_2}$ of 13.4~kHz and $g_{H}^{(2)}(0)=0.013\pm0.009$. By increasing the pump power, the raw heralded rate increases while $g_{H}^{(2)}(0)$ decreases. At pump power of 1.97~$\mu$W, we observe HSPS with a raw heralded rate of 104.8~kHz and $g_{H}^{(2)}(0)=0.196\pm0.012$. At the highest pump power of 3.3~$\mu$W in our experiment, we observe HSPS with a raw heralded rate of 170.7~kHz and $g_{H}^{(2)}(0)=0.271\pm0.011$, which is still well below the classical threshold of 0.5.  

{\noindent\bf\em Conclusion.}---We demonstrate a MPM-enabled SPDC quantum photon source on a dual-layer LNOI waveguide. The MPM-enabled SPDC is able to generate photon pairs with high counter rate of 41.77~GHz/mW and high signal-to-noise ratio of up to 58298. Moreover, the SPDC source is able to generate HSPS with raw count rate higher than 100~kHz and high quality of $g_{H}^{(2)}(0)<0.2$, which is comparable to the state-of-art PPLN waveguide~\cite{zhao2020_Phys.Rev.Lett._High}. To give a comparison, we summarize the SPDC sources with MPM and QPM on LNOI platform as shown in Table~\ref{Tb:spdc}. Compared to the dual-layer LNOI obtained by poling process~\cite{shi2024efficient}, our stacking/bonding process is promising for applications where large size dual-layer LNOI is required. The dual-layer LNOI waveguide enables efficient frequency conversion with on-chip device. Our MPM-enabled SPDC source is of great potential for integrated quantum photonic applications.

\begin{table*}
\caption{A summary of SPDC sources with MPM and QPM on LNOI platform. N/A: Cannot be inferred from the reported data.}\label{Tb:spdc}
\begin{threeparttable}
\begin{tabular}{p{2.5cm}<{\centering}p{2.5cm}<{\centering}p{2.5cm}<{\centering}p{2.5cm}<{\centering}p{5cm}<{\centering}}
    \hline\hline
    &Type &Brightness &CAR &$g_{\text{H}}^{(2)}$(0)~(heralded rate)\\  
     \hline
    This work
    & MPM-LNOI
    & 41.77 GHz/mW
    & 58298
    & 0.196~(105kHz)     \\
    \hline
    Shi~\emph{et al.}~\cite{shi2024efficient}
    & MPM-LNOI
    & 3 MHz/mW$^2$/nm\tnote{b}
    & 2043
    & 0.008 (29kHz\tnote{e} )\\
    \hline  
    Zhao~\emph{et al.}~\cite{zhao2020_Phys.Rev.Lett._High}
    & QPM-LNOI
    & 56 MHz/mW/nm\tnote{c}
    & 67224
    & 0.183~(91kHz)  \\
    \hline    
    Javid~\emph{et al.}~\cite{javid2021_Phys.Rev.Lett._Ultrabroadband}
    & QPM-LNOI
    & 13 GHz/mW
    & 20248
    & N/A\\
    \hline
    Xue~\emph{et al.}~\cite{xue2021_Phys.Rev.Applied_Ultrabright}
    & QPM-LNOI
    & 279 GHz/mW
    & 599
    & N/A\\
    \hline
    Henry~\emph{et al.}~\cite{henry2023_Opt.Express_Correlated}
    & QPM-SiN rib\tnote{a} 
    & 4.8 GHz/mW\tnote{d}
    & 7680
    & 0.04~(48kHz) \\  
    \hline
    Zhang~\emph{et al.}~\cite{zhang2023_Optica_Scalable}
    & QPM-LNOI
    & 178 MHz/mW
    & 8136
    & N/A\\
    \hline
    Fang~\emph{et al.}~\cite{fang2024_Opt.Express_Efficient}
    & QPM-LNOI
    & 11.7 GHz/mW
    & 16262
    & 0.198~(27.9kHz) \\
    \hline
    \hline
\end{tabular} 
    \begin{tablenotes}
    \footnotesize
    \item[a] Hybrid SiN-LNOI waveguide, where the 400~nm SiN is etched by 350~nm.
    \item[b] Data from a cascaded SHG and SPDC process and the waveguide using telecom cw pump.
    \item[c] Measured 45~MHz/mW after the filtering of 100~GHz.
    \item[d] Calculated from the PGR of 23~MHz/mW with 100~GHz filtering and the bandwidth of 21~THz.
    \item[e] Estimated from the Figure 3 (c) in \cite{shi2024efficient}.
    \end{tablenotes}
    \end{threeparttable}
\end{table*}

\begin{acknowledgements} %delete if not applicable))
{\noindent\bf\em Acknowledgments.}---This work is supported by the National Key R\&D Program of China (Grants No.~2019YFA0308200 and No.~2019YFA0705000), the National Natural Science Foundation of China (Grant No.~92065112), Shandong Provincial Natural Science Foundation (Grant No.~ZR2023LLZ005), Taishan Scholar of Shandong Province (Grants No.~tsqn202103013 and No.~tspd20210303), Shenzhen Fundamental Research Program (Grant No.~JCYJ20220530141013029) and the Higher Education Discipline Innovation Project (``111") (Grant No. B13029). 
\end{acknowledgements}
%#########################################################################################################################################################################################

%\bibliographystyle{MSP}
\bibliographystyle{apsrev4-2}
\bibliography{LNMPM}
%\begin{thebibliography}{10}
%\providecommand{\url}[1]{\texttt{#1}}
%\providecommand{\urlprefix}{URL }
%
%
%
%\end{thebibliography}

\end{document}